\documentclass[prb,amsmath,amssymb,twocolumn,superscriptaddress,showpacs]{revtex4-1}
\usepackage{bm}
\usepackage{amssymb}
\usepackage{colordvi}
\usepackage{graphicx}
\usepackage{color}
\usepackage{hyperref}
\usepackage{ulem}

\newcommand{\beq}{\begin{equation}}
\newcommand{\eeq}{\end{equation}}
\newcommand{\bea}{\begin{eqnarray}}
\newcommand{\eea}{\end{eqnarray}}

\begin{document}

\title{Photonic Chern insulators made of gyromagnetic hyperbolic metamaterials}

\author{Ruei-Cheng Shiu}
\affiliation{Department of Physics and Center for Theoretical Physics, National Taiwan University, Taipei 10617, Taiwan}
\author{Hsun-Chi Chan}
\affiliation{Department of Physics and Center for Theoretical Physics, National Taiwan University, Taipei 10617, Taiwan}
\author{Hai-Xiao Wang}
\affiliation{Department of Physics and Center for Theoretical Physics, National Taiwan University, Taipei 10617, Taiwan}
\affiliation{Physics Division, National Center for Theoretical Sciences, Hsinchu 30013, Taiwan}
\author{Guang-Yu Guo}\email{gyguo@phys.ntu.edu.tw}
\affiliation{Department of Physics and Center for Theoretical Physics, National Taiwan University, Taipei 10617, Taiwan}
\affiliation{Physics Division, National Center for Theoretical Sciences, Hsinchu 30013, Taiwan}

\date{\today}

\begin{abstract}
Controlling light propagation using artificial photonic crystals and electromagnetic metamaterials is 
an important topic in the vibrant field of photonics. Notably, chiral edge states on
the surface or at the interface of photonic Chern insulators can be used to make reflection-free waveguides. 
Here, by both theoretical analysis and electromagnetic simulations,
we demonstrate that gyromagnetic hyperbolic metamaterials (GHM)
are photonic Chern insulators with superior properties.
As a novel mechanism, the simultaneous occurrence of the hyperbolic and gyromagnetic effects in these metamaterials
is shown to open the large topological band gaps with gap Chern number of one.  
Importantly, unlike many other photonic Chern insulators, the GHM Chern insulators possess
non-radiative chiral edge modes on their surfaces, and thus allow to fabricate unidirectional waveguides
without cladding metals which generally incurr considerable Ohmic loss.
Furthermore, the photonic edge states in the proposed Chern insulators are robust against disorder 
on a wide range of length scales, in strong contrast to crystalline topological insulators, 
and the light flow direction on the surface of the Chern insulators can be easily 
flipped by switching the direction of an applied magnetic field. Fascinatingly, we
find that negative refraction of the topological surface wave occurs at the boundary between
the GHMs with the opposite signs of gyromagnetic parameters. Finally, we show that 
compared with other photonic topological materials such as chiral hyperbolic materials, the present 
GHM Chern insulators can be much easier to fabricate.

\end{abstract}

\maketitle

\section{Introduction}
Control over the propagation of light using artificial photonic crystals\cite{PhC_John,PhC_Yablonovitch} 
and electromagnetic metamaterials\cite{MetaSmith} has received enormous attention in recent decades mainly 
because of its importance for many applications in the vibrant field of photonics. For example, metamaterials 
such as left-handed media\cite{Veselago68,Chan18} have shown promising potential for novel 
technologies\cite{Pendry00,Klimov12}. In recent decades, great progress in this field 
has been often made by taking advantages of analogies with electronic systems in solid state physics. 
For example, the concept of a photonic band gap material\cite{PhC_John,PhC_Yablonovitch}, a man-made system 
with a periodic dielectric function, was inspired by the electronic Bloch states in a crystalline semiconductor.

More recently, there have been growing interests in using topological photonic materials\cite{Lu2014} 
to manipulate the flow of light, again inspired by the recent developments of electronic topological 
materials \cite{Hasan2010,Qi2011,hmwengreview}. In particular, the electronic quantum anomalous Hall (QAH) phase 
is a two-dimensional (2D) bulk ferromagnetic insulator with a nonzero Chern number in the presence of spin-orbit 
coupling (SOC) but in the absence of applied magnetic fields\cite{Haldane1988,QAHexp}. Its associated metallic 
chiral edge states in this Chern insulator carry dissipationless unidirectional electric current. Haldane and Raghu 
recently proposed\cite{photon_QAH_Haldane} to construct analogs of this intriguing QAH in photonic crystals made 
of time-reversal symmetry (TRS) breaking materials to realize unidirectional optical waveguides. Subsequently, 
these topological electromagnetic states in a number of gyromagnetic photonic crystals with broken TRS were 
further proposed \cite{QAH_Wangzhen_theory,ochiai2009photonic,QAH_poo2011,zeromode,QAH_largeChern,Dirac_mirrorsym,Chan2018} 
and observed\cite{QAH_Wangzhen_exp,QAH_poo2013,QAH_largeChern_exp,He2016}. Interestingly, photonic analog of 
electronic quantum spin Hall effect in 2D topological insulators with TRS\cite{Hasan2010,Qi2011} were also 
observed in bi-anisotropic photonic crystals\cite{Khanikaev2013,z2new,z2}. 

Nevertheless, investigations of photonic unidirectional edge modes have mostly been limited to topological 
photonic crystals periodic on the scale of the operational wavelength, and this considerably restricts the 
applications of topological photonic materials. Very recently, Gao et. al.\cite{chiralhypermeta} theoretically 
demonstrated topological photonic phase in chiral hyperbolic metamaterials (CHM) made of continuous TRS media 
with photonic edge states robust against disorder on all length scales\cite{chiralhypermeta}. 
In hyperbolic metamaterials\cite{HMM13}, 
which are plasmonic metamaterials, equi-frequency surfaces (EFSs) of transverse electric-field (TE) and 
transverse magnetic-field (TM) modes are degenerate on the high-symmetry points in the momentum space [see Fig. 1(b)]. 
When the bi-anisotropic property (chirality) is introduced in hyperbolic metamaterials with TRS, 
which then become CHMs, the degeneracies are broken due to the coupling between TE and TM modes\cite{Khanikaev2013} 
and consequently, a nontrivial band gap is opened\cite{chiralhypermeta}. The nontrivial topology of the CHM 
results from the nonzero Berry curvature due to the chirality (equivalent to the SOC in electronic topological 
insulators) and broken spatial inversion symmetry in continuous medium\cite{continuousmediachern}. Furthermore, 
the design principle was based on the concept of  a Floquet topological insulator (FTI)\cite{floquetrechtsman2013} 
where by treating the distance $z$ propagated by a waveguide mode as a timelike coordinate, e.g., 
a honeycomb lattice of helical waveguides behaves like a FTI.
 
In this work, as a novel mechanism for controlling light flow, we introduce the photonic Chern insulators 
made of continuous gyromagnetic hyperbolic metamaterials (GHM) with the band gap opened by the TRS-breaking 
gyromagnetic effect [Figs. 1(b) and (c)]. The non-trivial topology is demonstrated by the calculated Berry 
curvature and nonzero Chern number due to the broken TRS. The unidirectional backscattering-immune non-radiative 
edge modes at the interface between the GHM and vacuum are uncovered by the finite-element electromagnetic simulations. 
As in the photonic quantum spin-Hall insulators (QSHI) made of the CHMs\cite{chiralhypermeta}, the photonic 
edge states in our Chern insulators are robust against disorder on all length scales. In contrast to the 
QSHIs\cite{chiralhypermeta}, however, the light flow direction on the surface of our Chern insulators 
can be easily flipped by switching the direction of applied magnetic fields. Furthermore, our Chern insulators 
made of the GHM can be easily fabricated. 

\section{Gyromagnetic hyperbolic metamaterial}


We consider a GHM as a hyperbolic metamaterial\cite{HMM13} with the gyromagnetic response 
which is described by the constitutive relation,
\beq
\left( \begin{matrix}
\mathbf{D}  \\
\mathbf{B}  \\
\end{matrix} \right)=\left( \begin{matrix}
{{\varepsilon_0}}\hat{\varepsilon} & 0  \\
0 & {{\mu_0}}\hat{\mu}  \\
\end{matrix} \right)\left( \begin{matrix}
\mathbf{E} \\
\mathbf{H}  
\end{matrix} \right),\\
\eeq
where 
\beq
\hat{\varepsilon}=\left(
\begin{array}{ccc}
	\epsilon_{xx}  & 0 & 0 \\
	0 & \epsilon_{yy} & 0 \\
	0 & 0 & \epsilon_{zz} \\
\end{array}
\right),  
\hat{\mu }=
\begin{pmatrix}
	\mu_{xx} & -i\gamma & 0 \\
	i\gamma & \mu_{yy} & 0 \\
	0 & 0 & \mu_{zz}
\end{pmatrix}
\eeq
are the relative permittivity and permeability tensors, respectively, and $\gamma$ is the gyromagnetic parameter, 
representing the degree of TRS breaking upon application of a magnetic field in the $z$-direction (gyromagnetic effect).
For simplicity, let us set $\mu_{xx}=\mu_{yy}=\mu_{zz}=\mu=1$, $\epsilon_{xx}=\epsilon_{yy}=\alpha$ and
$\epsilon_{zz}=\beta$ such that EFS dispersion are isotropic in $X-Y$ plane. 
The propagation behavior in the gyromagnetic hyperbolic metamaterials can be described by the wave equation 
of the electric field $\mathbf{E}=(E_x,E_y,E_z)^T$ as 
\beq 
\label{eq_wave}
\mathbf{k}\times \hat{\mu}_r^{-1} \mathbf{k}\times\mathbf{E}+k_0^2 \hat{\epsilon_r} \mathbf{E}=0,
\eeq 
where $\mathbf{k}=(k_x,k_x,k_z)$ is the wave vector, and $k_0=\omega/c$ is the wave number in vacuum. The left 
side of Eq. \eqref{eq_wave} can be rewritten in the matrix form as
\beq 
\label{eq_waveMatrix}
\left(
\begin{array}{ccc}
\alpha k_0^2 - \frac{k_y^2}{\mu}-Ak_z^2  & \frac{k_xk_y}{\mu}+Bk_z^2 & Ak_xk_z-Bk_yk_z \\
\frac{k_xk_y}{\mu}-Bk_z^2 & \alpha k_0^2 - \frac{k_x^2}{\mu} - Ak_z^2 & Ak_yk_z+Bk_xk_z\\
Ak_xk_z+Bk_yk_z & Ak_yk_z-Bk_xk_z & \beta k_0^2-Ak_t^2\\
\end{array}
\right)
\eeq 
where $k_x^2+k_y^2=k_t^2$, $A=\frac{\mu}{\mu^2-\gamma^2}$, $B=\frac{\gamma}{\mu^2-\gamma^2}$. The nontrivial 
solutions of $\mathbf{E}=(E_x,E_y,E_z)^T$ exist when the determinant of Eq.~\ref{eq_waveMatrix} equals to zero, 
resulting in the eigen equation of
\beq
\label{eq_Det0}
\alpha ^2 \beta  k_0^2 \mu-(\alpha A\mu+\beta-A)\alpha k_t^2
+(A-2\alpha \mu A)\beta k_z^2=0.
\eeq
 
Starting with an isotropic optical medium with $\epsilon_{xx}=\epsilon_{zz}=2$ and $\gamma=0$, one has two routes 
to arrive at a GHM, as illustrated in Fig. 1. The EFS dispersion of the isotropic medium is a perfect sphere, 
as shown in Fig. 1(a). Furthermore, the eigenstates consisting of transverse electric mode TE ($E_x=E_y=0$, $H_z\neq0$) 
and transverse magnetic modes TM ($H_x=H_y=0$, $E_z\neq0$), are degenerate and their dispersion are identical 
and on the same sphere. When  the isotropic medium is transformed to a hyperbolic metamaterial 
($\epsilon_{xx}=2$, $\epsilon_{zz}=-1$, and $\gamma=0$) by varying $\epsilon_{zz}$ from 2 to -1, the EFS sphere 
of the TM mode splits and becomes two parabola along the $k_z$-axis while that of the TE mode remains spherical, 
as depicted in Fig. 1(b) [see Fig. 2(a, b) for polarization analysis]. This is because TM "sees" $\epsilon_{xx}=\epsilon_{yy}=2$ 
only while, in contrast, TM "sees" $\epsilon_{x-y}=2$ and =$\epsilon_{zz}=-1$. Interestingly, the two TM parabola 
touch the TE sphere at $k_z =\pm 1.0k_0$, respectively. When the gyromagnetic effect is further introduced into 
the system by making $\gamma$ nonzero (e.g., $\gamma=0.8$), the system becomes a GHM. As a result, a band gap 
is opened at both degenerate points $k_z =\pm1.0k_0$ [see Fig. 1(d)]. Interestingly, since the gyromagnetic effect breaks TRS, 
the eigenstates near $\pm 1.0k_0$ become circularly (elliptically) polarized, while eigenstates on the upper and 
lower hyperbolic bands far away from the singular points remain TM-polarized [see Fig. 2(c,d)]. 
 
\begin{figure}[htbp]
\includegraphics[width=3.2in]{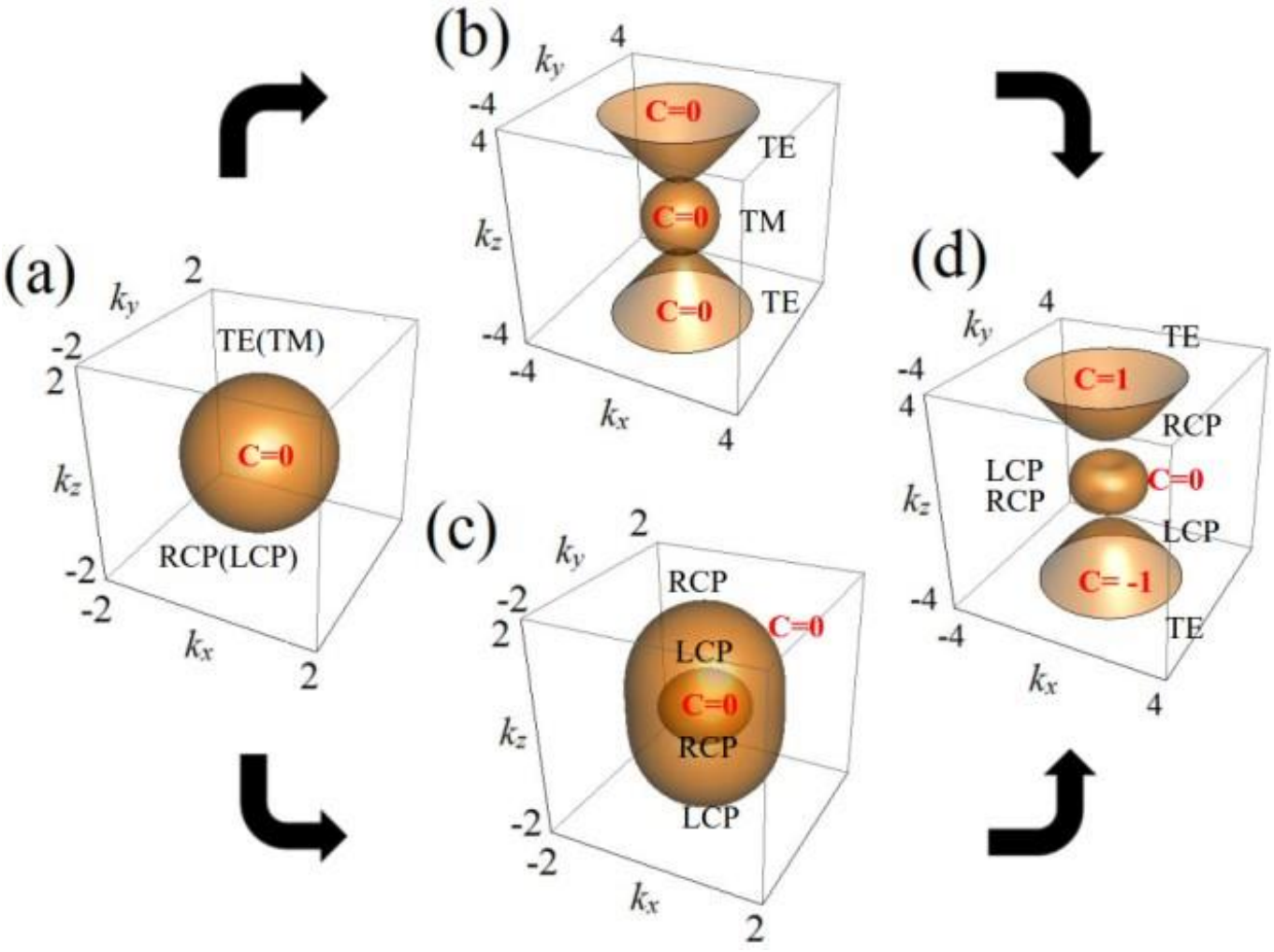}
\caption{EFS evolution from (a) an isotropic medium ($\epsilon_{xx}=\epsilon_{zz}=2$ and $\gamma=0$) to (d) 
a GHM ($\epsilon_{xx}=2$,$\epsilon_{zz}=-1$, and $\gamma=0.8$) via either (b) a hyperbolic metamaterial 
($\epsilon_{xx}=2$,$\epsilon_{zz}=-1$, and $\gamma=0$) or (c) a gyromagnetic medium ($\epsilon_{xx}=\epsilon_{zz}=2$ 
and $\gamma=0.8$). TE (TM) and RCP (LCP) denote TE (TM) polarization and right-handed (left-handed) circular 
polarization, respectively.} 
\end{figure}

Alternatively, one can go from an isotropic medium first to a gyromagnetic medium (e.g., $\epsilon_{xx}=\epsilon_{zz}=2$ 
and $\gamma=0.8$) [see Fig. 1(c)] and then to the GHM. In a gyromagnetic medium, because of broken TRS, TE and TM 
modes are coupled, resulting in the separation of the degenerate EFS spheres into two ellipsoids [Fig. 1(c)]. 
Also eigenstates becomes elliptically polarized [see Fig. 2(e,f)]. In particular, the eigenstates on the top 
and bottom surfaces of the ellipsoids become nearly fully circularly polarized, while those in the vicinity 
of the $k_z=0$ plane remain almost purely either (inner ellipsoid) TE or (outer ellipsoid) TM polarization. 
Finally, when $\epsilon_{zz}$ is tunned to a negative value (e.g. $\epsilon_{zz}=-1$), the system becomes a GHM 
and the outer ellisoid splits to form two open parabolas, as shown in Fig. 1(d).

\begin{figure}[htbp]	
\includegraphics[width=3.2in]{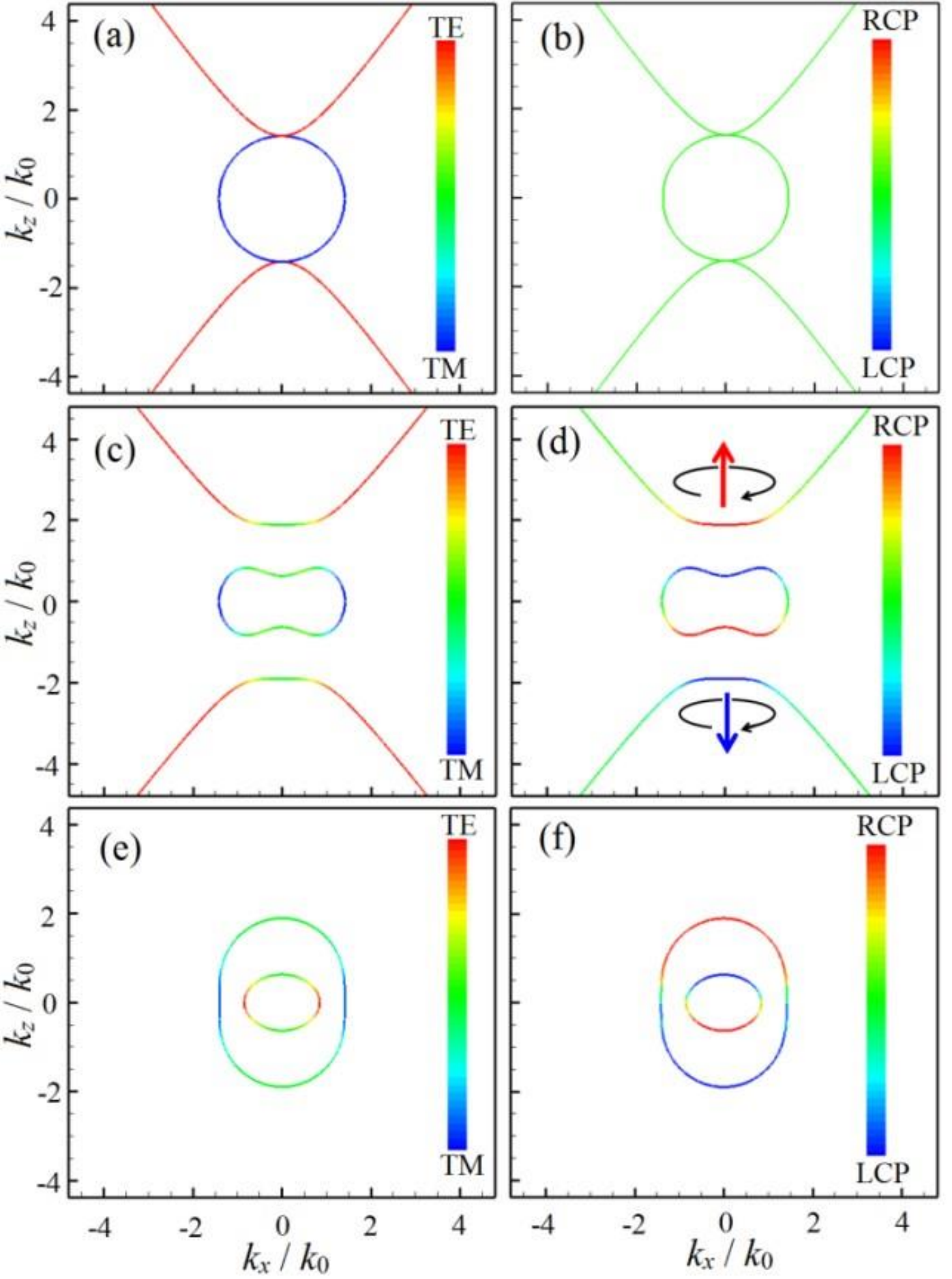}
\caption{Polarization analysis of eigenstates. Calculated polarization of EFS dispersion for (a, b) the hyperbolic
metamaterial, (c, d) the GHM (c, d) and (e, f) the gyromagnetic medium. RCP (LCP) and TE (TM) denote
right-handed (left-handed) circular polarization and TE (TM) polarization, respectively.}
\end{figure}

\section{Topological phase transition}
 
To investigate the topological property of the EFS dispersions of all four kinds of optical media and metamaterials 
and also to examine the topological nature of the band gaps near $k_z=\pm k_0$, 
we calculate the Berry phase $\Phi =\iint{\mathbf{F}(\mathbf{k})\cdot ds}$ and hence the Chern number $\mathcal{C}=\tfrac{1}{2\pi}\Phi$ 
of all EFS surfaces.\cite{fukui2005chern,Chan2018} Following our previous work\cite{Chan2018},
we adopt the efficient numerical algorithm reported in Ref. [\onlinecite{fukui2005chern}] to 
evaluate the Berry curvature $\mathbf{F}(\mathbf{k})$. Note that in principle, the Berry phase and the Chern number are
well defined only for a closed surface. Fortunately, our test calculations show that the Berry curvature $\mathbf{F}(\mathbf{k})$
is negligibly small when the radial wave vector $k_\rho=\sqrt{k_x^2+k_y^2}>15$.
Therefore, although the hyperbolic shaped EFS surfaces are open (Figs. 1 and 2), we find that the calculated Chern numbers would converge 
well to integers as long as the surface integration is carried out from the $k_0$ up to the radial wave vector $k_\rho=\sqrt{k_x^2+k_y^2}$
being larger than $15$. The Chern numbers calculated in this way are shown in Fig. 1, and the Berry curvature distributions 
are displayed in Fig.~3. We notice that the Berry curvature has the following symmetry properties: 
(a) $\mathbf{F}(\mathbf{k})=\mathbf{F}(-\mathbf{k})$ if the system has the spatial inversion symmetry (IS) and 
(b) $\mathbf{F}(\mathbf{k})=\mathbf{-F}(-\mathbf{k})$ if the system has the TRS symmetry. Therefore, the Berry curvature 
is identically zero and hence the Chern number is zero for both isotropic medium and hyperbolic metamaterial because they have 
both IS and TRS [See Figs.~1(a) and 1(b)].

\begin{figure}[htbp]
\includegraphics[width=3.2in]{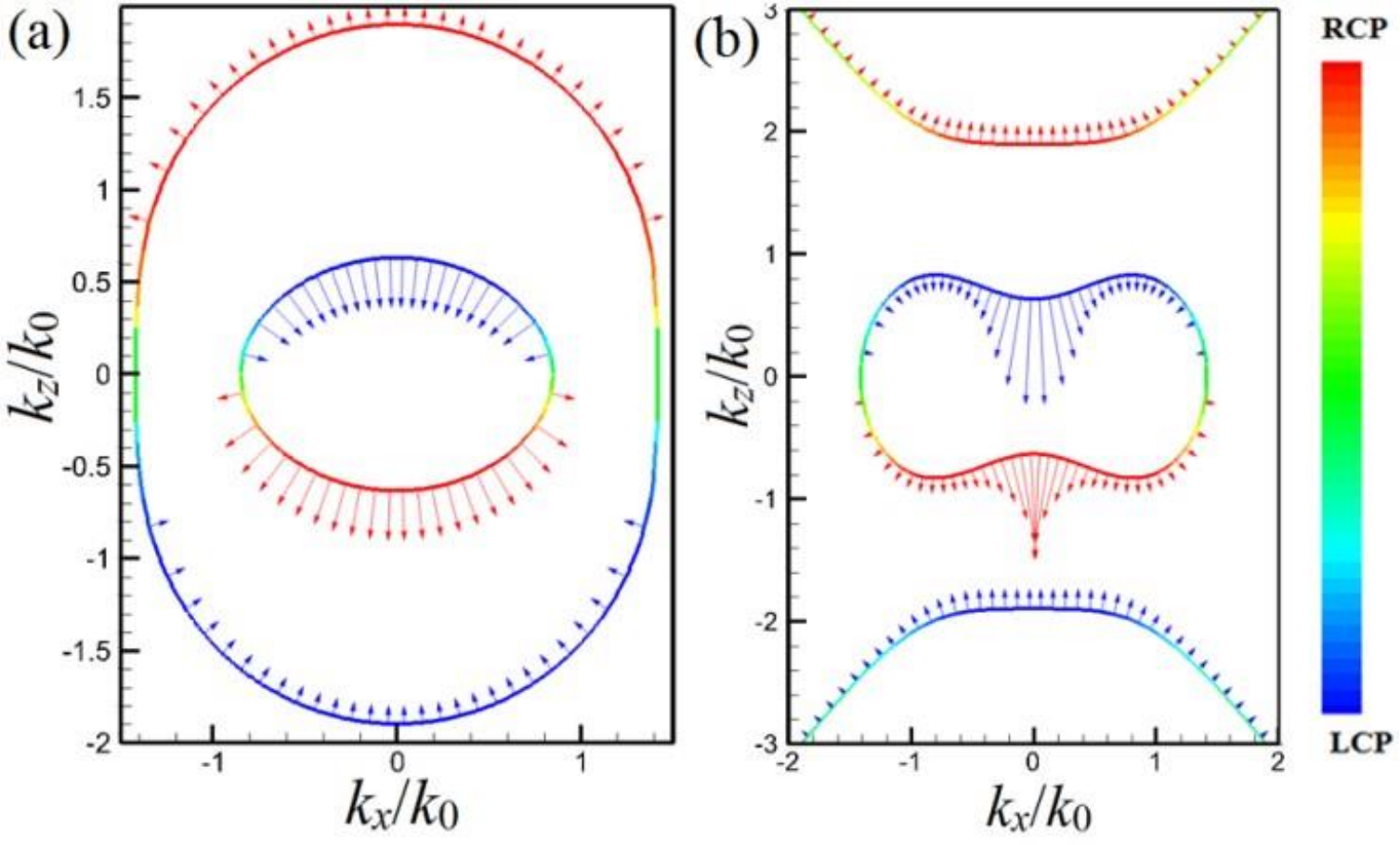}
\caption{Berry curvature distribution on EFSs for the gyromagnetic hyperbolic metamaterials with
(a) $\alpha=\beta=2$, and $\gamma=0.8$, (b) $\alpha=2$, $\beta=-1$, and $\gamma=1.2$.}
\end{figure}
 
In contrast, the Berry curvatures become nonzero in both gyromagnetic medium and GHM since their TRS is broken, 
as shown in Fig.~3. Nevertheless, in the gyromagnetic medium, the Berry curvatures of the eigenstates on each EFS ellipsoid 
form an odd function of $k_z$ with respect to the out-surface element [Fig. 3(a)]. Consequently, the sum of 
the Berry curvatures on the EFS surface of each ellipsoid (i.e., the Berry phase and the Chern number) remains zero [Fig. 1(c)]. 
This is also true for the inner ellipsoid in the GHM [Fig. 3(b)] and hence the Chern number of the inner EFS is zero [Fig. 1(d)]. 
However, the outer ellipsoid in the GHM now splits and hence transforms into two open parabolic surfaces with the Berry curvatures 
pointing to the positive $z$-direction. Although the Berry curvatures on the upper and lower surfaces still form 
an odd function of $k_z$, they are on the different surfaces. Resultantly, the upper and lower open surfaces each 
acquires a Chern number of -1. Therefore, the phase transformation from the hyperbolic to gyromagnetic hyperbolic 
due to the gyromagnetic effect is a topological one, and the band gaps centered at $\pm k_0$ are topological nontrivial. 

\section{Nonradiative one-way photonic edge mode}
Next, we follow the method suggested in Ref. [\onlinecite{Dya88}] to calculate analytically the surface band between the dielectric and 
anisotropic material. Let the interface between air ($y>0$) and GHM ($y<0$) be on the $xz$ plane ($y=0$). The wave vector in 
GHM normal to the interface can then be obtained by solving Eq. \eqref{eq_Det0}

\begin{widetext}
\begin{equation} 
\label{eq_ky}
k_y^{GHM} = \pm \left( \frac{\alpha  k_0^2 \left(\alpha  \mu ^2+\beta  \left(\mu
	^2-\kappa ^2\right)\right)-2 \alpha  \mu  k_x^2-\alpha 
	\mu  k_z^2-\beta  \mu  k_z^2 \pm Y}{2 \alpha  \mu }\right) ^{0.5}.
\end{equation}
where
\begin{equation}
Y \equiv \sqrt{\alpha ^2 k_0^4 \left(\alpha  \mu ^2+\beta 
	\left(\kappa ^2-\mu ^2\right)\right)^2+2 \alpha  k_0^2
	\mu  k_z^2 \left(-\alpha ^2 \mu ^2+\alpha  \beta 
	\left(\kappa ^2+2 \mu ^2\right)+\beta ^2 \left(\kappa
	^2-\mu ^2\right)\right)+\mu ^2 (\alpha -\beta )^2 k_z^4}
\end{equation}

Then two linear independent eigen fields in GHM $\vec{E}_i(k_x,k_y,k_z)$ ($i=3,4$) can be obtained by solving the null space 
of Eq. \eqref{eq_waveMatrix} with $k_y=k_y(k_x,k_z)$ from Eq. \eqref{eq_ky}. Once $\vec{E}_i$ are obtained, $\vec{H}_i$ 
can also be obtained via Faraday's law $\vec{H}_i=\frac{1}{Z_0}\vec{k} \times \vec{E}_i$. Likewise, two orthogonal eigen fields 
in the air can be expressed as 
\begin{equation}
\vec{E}_1=\left(
\begin{array}{c}
-k_z \\
0 \\
k_x \\
\end{array}
\right), \quad
\vec{E}_2=\left(
\begin{array}{c}
-k_y \\
k_x \\
0 \\
\end{array}
\right), \quad
\vec{H}_1=\frac{1}{Z_0}\left(
\begin{array}{c}
k_x k_y \\
-k_x^2-k_z^2 \\
k_y k_z \\
\end{array}
\right), \quad
\vec{H}_2=\frac{1}{Z_0}\left(
\begin{array}{c}
-k_x k_z \\
-k_y k_z \\
k_x^2+k_y^2 \\
\end{array}
\right),
\end{equation}
where $Z_0=\sqrt{\mu_0/\epsilon_0}$ is the vacuum impedance, and $k_y=i\sqrt {k_x^2 + k_z^2 - k_0^2}$. Since the tangential 
components of electromagnetic fields should be continuous across the interface, we arrive at $Det(D)=0$ with
\begin{equation}
D(k_x,k_z)=
\left(
\begin{array}{cccc}
-k_z & -k_y & E_{3x} & E_{4x} \\
k_x & 0 & E_{3z} & E_{4z} \\
k_x k_y & -k_x k_z & H_{3x} & H_{4x} \\
k_y k_z & k_x^2+k_y^2 & H_{3z} & H_{4z} \\
\end{array}
\right).
\end{equation}
In order to obtain a surface mode, it is important to choose the range of the imaginary part of $k_y$ ($k_y^{GHM}$) 
in the vacuum (GHM) to be positive (negative). In other words, the fields in the air and GHM must decay exponentially 
along the vertical direction of the interface.  
\end{widetext}

According to the bulk-edge correspondence, the gap Chern numbers of $\Delta C=-1$ implies that there is 
one propagating edge state in each band gap for each air-GHM interface. Furthermore, these edge states 
are chiral and topology-protected, i.e., they are reflection-free one-way edge states. To verify these 
amazing predictions, we solve the Maxwell’s equations for a slab made of the GHM, as illustrated in Fig.~4(a). 
The calculated EFS dispersions for this slab, displayed in Fig. 4(b), show that in addition to the bulk bands 
(black curves), there is indeed one edge state for both top (red curve) and bottom (blue curve) interface 
between air and the GHM slab in each band gap.

We further simulate light propagation in the air-GHM interface by putting a line source along the $z$-axis 
at $x = 0$. The results are displayed in Figs. 4(c)-4(f). Our simulations show clearly that light on 
each interface runs without radiation in either $+x$ or $–x$ direction but not both directions, i.e., 
light propagation on the edges is unidirectional. We notice that in Fig.~4(c), both edges at $k_z=1.6k_0$ 
carry light towards $–x$ direction. This is because the dispersions of both bulk and edge states are 
symmetric with respect to the mirror located at $k_z=0$. And the propagation along the $–x$ direction is 
caused by the negative gap Chern numbers ($\Delta C=-1$) of both band gaps. Moreover, Fig.~4(g) shows 
that light can overcome the obstacles on the surface, indicating light propagation is reflection-free 
due to the topological protection. Finally, the direction of light propagation depends on the sign 
of gyromagnetic parameter $\gamma$ [see Figs. 4(c) and 4(d)], demonstrating that one can control light 
transmission direction by switching the magnetization direction of the GHM slab.    

\begin{figure}
\includegraphics[width=3.2in]{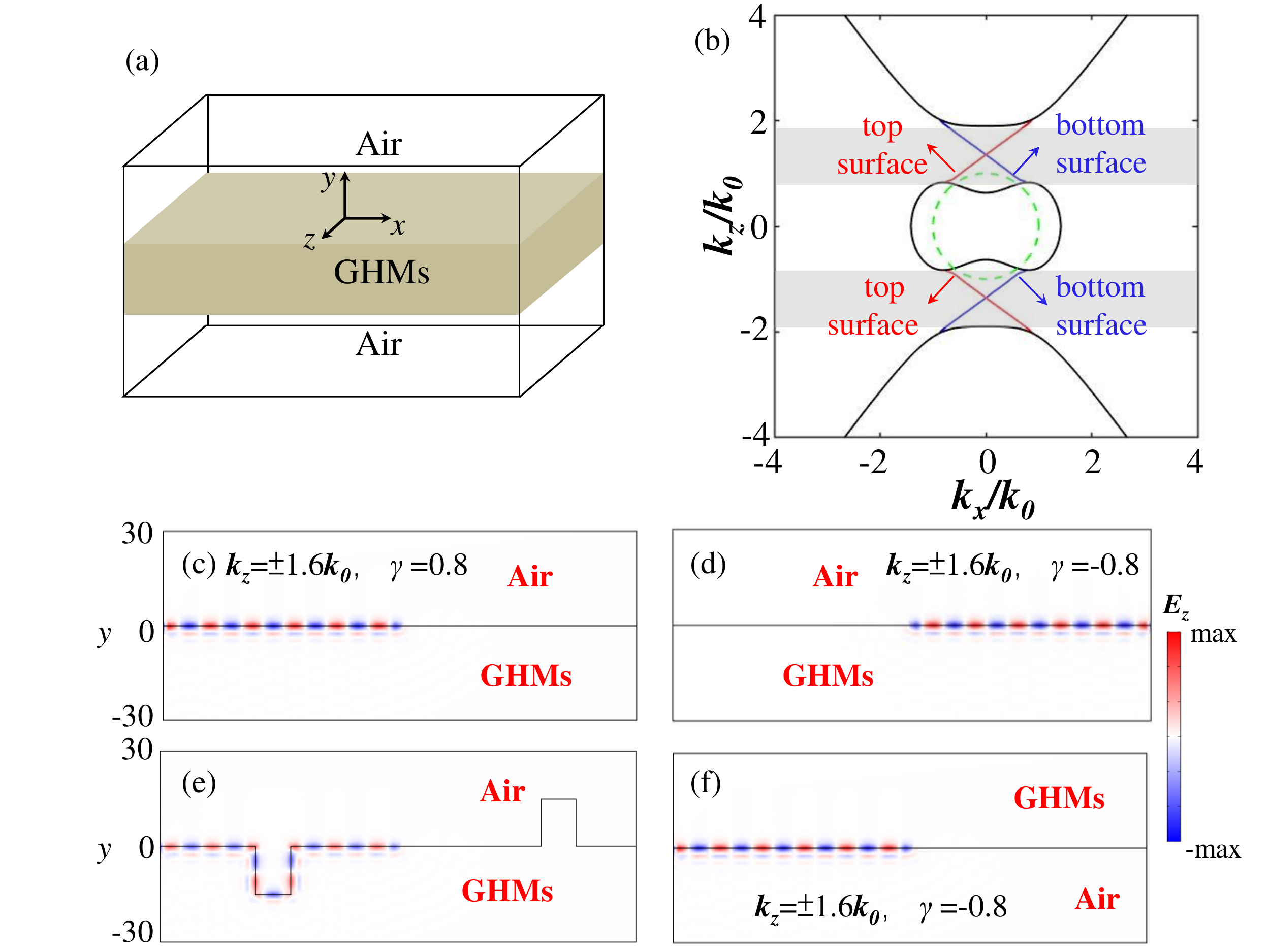}
\caption{One-way propagating edge states. (a) A GHM slab in air. (b) Calculated EFS dispersions for 
the GHM slab in air (a). Black, red, and blue lines denote dispersions of bulk state, top and bottom edge states, 
respectively. The dashed green circle denotes light cone. (c) and (d) Simulated propagations of light emitted 
by a line source at $x=0$ on the top surface at $k_z=1.6k_0$ with $\gamma=-0.8$ and $\gamma= +0.8$, respectively. 
(e) The same as in (c) except that the surface is now uneven, which does not stop light propagate along the negative $x$-direction.
(f) The same as in (d) but on the bottom surface. 
}
\end{figure}

\begin{figure}
\includegraphics[width=3.2in]{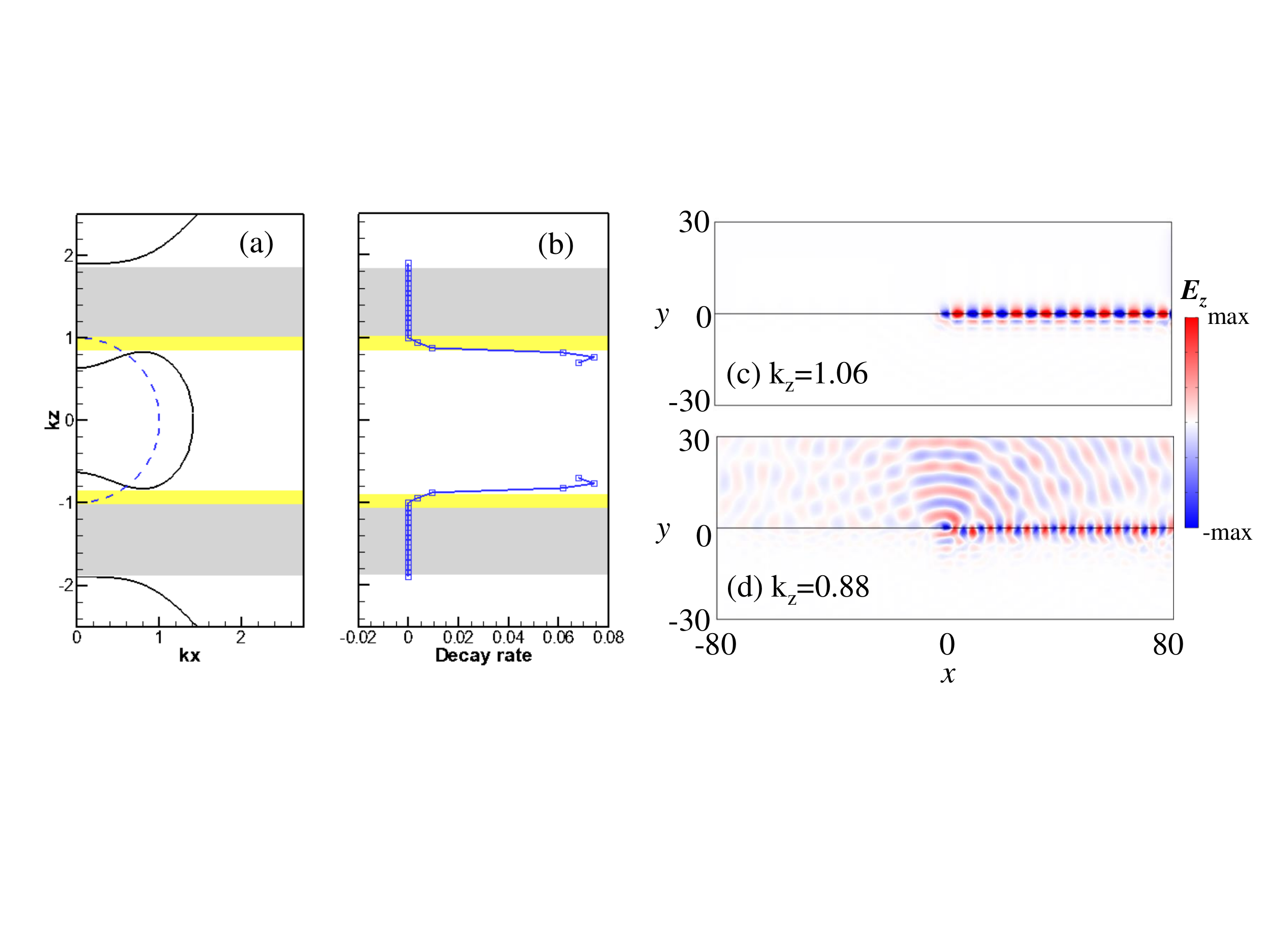}
\caption{(a) EFS dispersions and (b) decay rate of the electromagnetic waves for the GHM slab in the air. 
In (a), the solid black and dashed blue curves denote the dispersions of the bulk states
and the air, respectively. In (a) and (b), the yellow regions denote the overlap regions 
of the air dispersion and bulk gap regions, while the grey regions denote the complete gap regions.
(c) and (d) Electric field distributions of the edge states with $k_z=1.06$ and $k_z=0.88$, respectively.
Note that in (c), the electric field is completely confined to the edge of the GHM slab and 
hence there is no radiative energy loss, because the $k_z$ is outside the light EPS sphere.
In contrast, in (d), because the $k_z$ is within the light EPS sphere, the electromagnetic wave
of the edge state propagates strongly into the air, resulting in a large radiative energy loss.
}
\end{figure}

A GHM is based on a hyperbolic metamaterial, which is generally a
bilayer superlattice consisting of a metal layer and a dielectric layer stacked
along the normal direction (see next Section for details). Thus, the GHMs
possess the property of surface plasmon polaritons (SPPs). Compared with
the electromagnetic waves in the air, the SPPs are slower waves and hence
have larger wave vectors than that of light in the air. As a result, the edge
states on a GHM may inherit this larger wave vector propagation, as shown in
Fig. 5. Therefore, there may be no intersection between the air light line and
edge states, i.e., the edge states are out of the light line. Consequently, the
edge state would propagate on the interface without radiation [Fig. 5(c)] because the
edge states cannot couple to the electromagnetic waves in the air. In strong
contrast, most of photonic topological insulators are made of photonic
crystals and consequently, the one way edge states on the interfaces of these insulators
suffer from severe radiation loss and a metal wall serving as a cladding perfect
electric conductor or a perfect magnetic conductor has to be added to the
surrounding edges to stop radiation loss into the air \cite{QAH_Wangzhen_exp}. 
In practice, there is no lossless perfect conductor and even good metals such as copper
used as the cladding walls absorb the radiation significantly. This certainly hinders 
their applications. In contrast, in a GHM, the air serves as the good insulator for
stopping the radiation of the edge states into the air [Fig. 5(c)].

\section{Negative refraction of edge mode}
\begin{figure}
\includegraphics[width=3.2in]{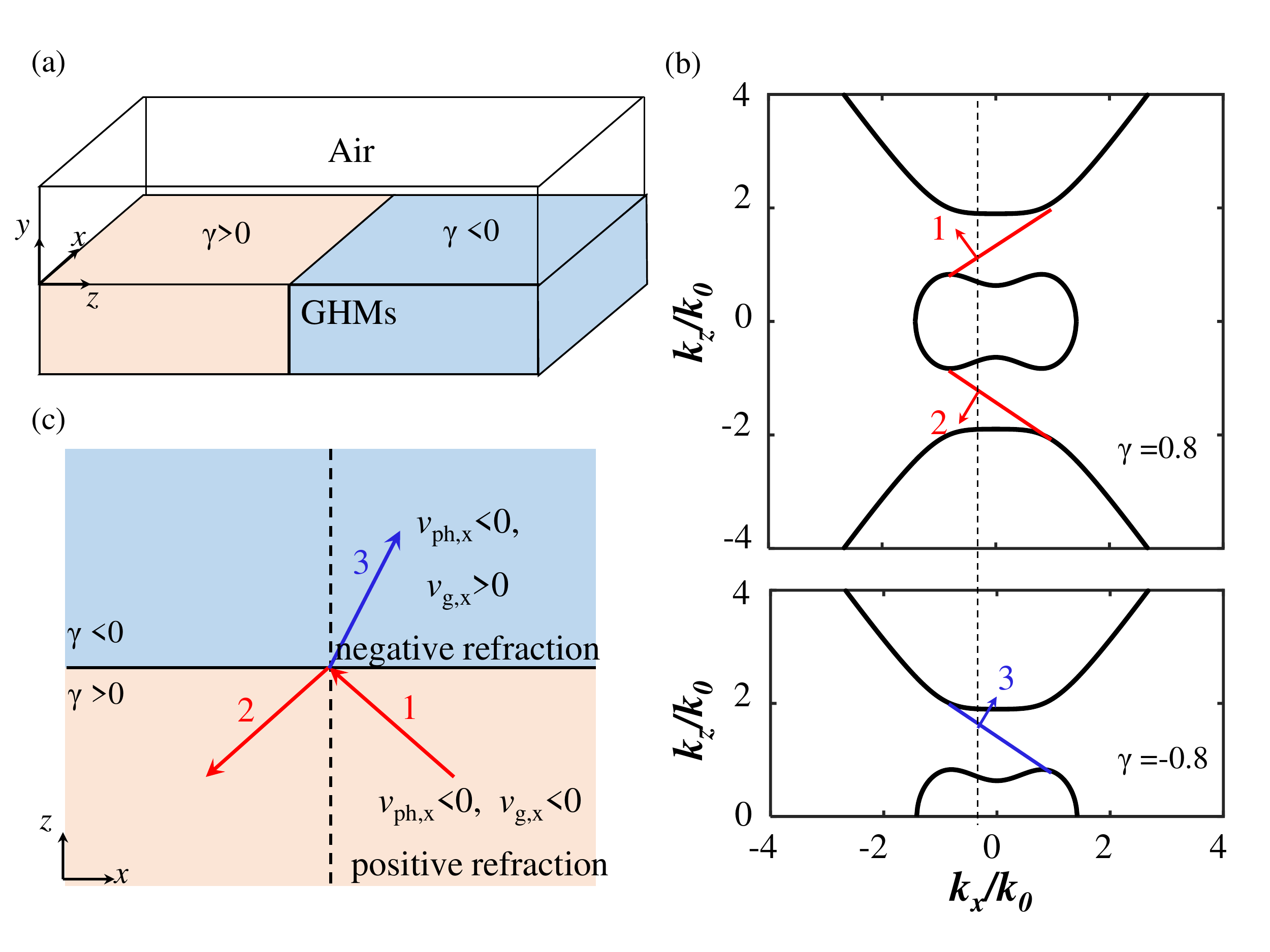}
\caption{\label{figNR}
Schematic illustration of the negative refraction of the TPSS of the GHM. 
(a) Side view and (b) the EFS of the system. (c) The incident,
reflected, and transmitted waves at the interface between the two different GHM insulators.
whose group velocities are denoted as  1, 2, and 3 in (b), respectively.
Here $v_{vg,x}$ and $v_{ph,x}$ denote the $x$ components of the group and phase velocities, respectively.}
\end{figure}

Another interesting property of the topological surface states of the GHM 
is negative refraction, similar to that of the CHM reported recently in Ref. [\onlinecite{Klimov18}]. 
As shown in Fig. 6(b), the $x$ component of the group velocity (energy flow) of all the surface states 
of the GHM with $\gamma=-0.8$ (the blue line in the bottom panel) is positive ($v_{v,x}>0$),
while the phase velocity ($v_{ph,x} = \omega/k_x$) changes sign across $k_z=0$. As a result, 
in the $k_x<0$ region the $x$ components of the group velocity and phase velocity have opposite signs 
($v_{g,x}>0$ and $v_{ph,x}<0$) [see Fig. 4(b)]. This is the signature 
of negative refraction \cite{HMM13,Klimov18}. On the contrary, the surface states on the GHM with $\gamma=0.8$
in the $k_x<0$ region have $v_{g,x}<0$ and $v_{ph,x}<0$ [see upper panel of Fig. 6(b)]. 
Following the method proposed in Ref. [\onlinecite{He18}], 
we show in Fig. \ref{figNR} that by drawing a vertical line in the EFS diagram (i.e., conservation of the momentum 
parallel to the interface), the group velocities of the three points intersecting the surface modes will 
give the propagation directions of the surface waves [Fig. \ref{figNR}(b)]. Consequently, as shown in Fig. \ref{figNR}(c), 
the incident (wave 1) and refracted (wave 3) surface waves lie on the same side of the normal line, indicating a negative refraction.
The existence of negative refraction of the topological surface states would lead to many interesting 
effects\cite{Veselago68,Pendry00,Chan18,He18,Fang05,Liu07,Smolyaninov07,Lin08} 
such as superlens effect\cite{Pendry00,Fang05,Liu07,Smolyaninov07} and suppression of reflected waves \cite{He18}.

\section{Realization of the proposed gyromagnetic hyperbolic metamaterials}

The hyperbolic metamaterials are a highly anisotropic material with real parts of the principal components 
of its permittivity tensor having opposite signs \cite{HMM13} and have been intensively investigated 
because of this unique property. Here we consider a bilayer superlattice composed of a metal layer with 
relative permittivity $\epsilon_m$ and thickness $d_m$ and a dielectric layer with relative 
permittivity$\epsilon_d$ and thickness $d_d$ stacked along the $z$-direction. The effective  dielectric constant 
of this metamaterial is given by 
\beq
\epsilon_{eff}=\left[
\begin{array}{ccc}
\epsilon_{x-y} & 0 & 0 \\
0 & \epsilon_{x-y} & 0 \\
0 & 0 & \epsilon_{z} \\
\end{array}
\right], 
\eeq
where
\begin{subequations}
\beq \epsilon_{x-y}=\frac{\epsilon_md_m+\epsilon_dd_d}{d_m+d_d}, \eeq 
\beq \frac{1}{\epsilon_{z}}=\frac{d_m/\epsilon_m+d_d/\epsilon_d}{d_m+d_d}. \eeq 
\end{subequations}

If the dielectric layer is replaced by a gyromagnetic medium layer, the superlattice becomes a gyromagnetic 
hyperbolic metamaterial with its relative permeability tensors
$\begin{bmatrix}
\mu_{g} & -i\gamma & 0 \\
i\gamma & \mu_{g} & 0 \\
0 & 0 & \mu_{g} \\
\end{bmatrix}$
and  
$\begin{bmatrix}
1 & 0 & 0 \\
0 & 1 & 0 \\
0 & 0 & 1 \\
\end{bmatrix}$
for the gyromagnetic and metal layers, respectively. The bilayer superlattice with relative 
permittivity $\epsilon_g= \epsilon_d$ and thickness $d_g= d_d$ for the gyromagnetic layer, is schematically shown in Fig. 7. 

\begin{figure}
\includegraphics[width=3.2in]{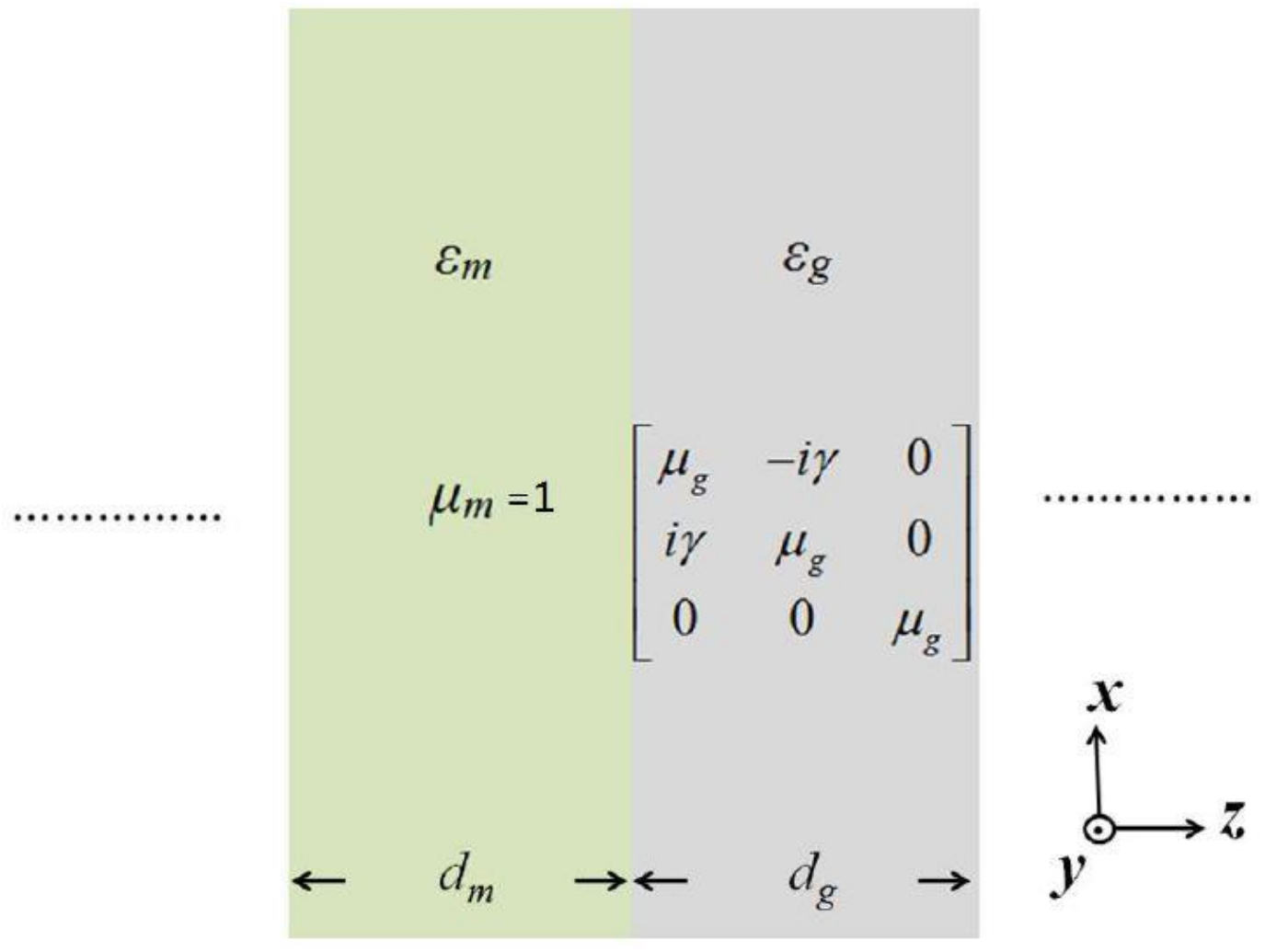}
\caption{Proposed gyromagnetic hyperbolic metamaterial as a bilayer
superlattice composed of a gyromagnetic slab and a metal slab.}
\end{figure}

For simplicity, let us write the fields in the metal as $(\mathbf{E}^g, \mathbf{H}^g)=(E_x^g, E_y^g, E_z^g, H_x^g, H_y^g, H_z^g)$ 
and in the gyromagnetic medium as $(\mathbf{E}^m, \mathbf{H}^m)=(E_x^m, E_y^m, E_z^m, H_x^m, H_y^m, H_z^m)$ . 
We assume that the wavelength is much larger than the thickness. Using the boundary conditions 
that $D_z$, $E_x$, $E_y$, $B_z$, $H_x$ and $H_y$ must be continuous across an interface, we obtain
$\epsilon_mE_z^m=\epsilon_gE_z^g$, $E_x^m=E_x^g=E_x$, $E_y^m=E_y^g=E_y$, $H_z^m=\mu_gH_z^g$, 
$H_x^m=H_x^g=H_x$, $H_y^m=H_y^g=H_y$. 
Consequently, the average $D$-field and $B$-field are 
\begin{subequations}
\beq D_x^{eff}=\frac{d_g\epsilon_g+d_m\epsilon_m}{d_g+d_m}E_x   \eeq
\beq D_y^{eff}=\frac{d_g\epsilon_g+d_m\epsilon_m}{d_g+d_m}E_y   \eeq
\beq D_z^{eff}=\epsilon_mE_z^m=\epsilon_gE_z^m                  \eeq
\beq B_x^{eff}=\frac{d_g \mu_g+d_m}{d_g+d_m}H_x -\frac{id_g\gamma}{d_g+d_m}H_y \eeq
\beq B_y^{eff}=\frac{d_g \mu_g+d_m}{d_g+d_m}H_y +\frac{id_g\gamma}{d_g+d_m}H_x  \eeq
\beq B_z^{eff}=\mu_0H_z^m=\mu_gH_z^m. \eeq
\end{subequations}
Therefore, the effective relative permittivity and permeability tensors are given by 
\begin{subequations}
\beq
\epsilon_{eff}=\left[
\begin{array}{ccc}
\frac{d_g\epsilon_g+d_m\epsilon_m}{d_g+d_m} & 0 & 0 \\
0 & \frac{d_g\epsilon_g+d_m\epsilon_m}{d_g+d_m} & 0 \\
0 & 0 & \frac{(d_g+d_m)\epsilon_g\epsilon_m}{d_g\epsilon_m+d_m\epsilon_g} \\
\end{array}
\right], 
\eeq

\beq
\mu_{eff}=\left[
\begin{array}{ccc}
\frac{d_g\mu_g+d_m}{d_g+d_m} & \frac{-id_g\gamma}{d_g+d_m} & 0 \\
\frac{id_g\gamma}{d_g+d_m} & \frac{d_g\mu_g+d_m}{d_g+d_m} & 0 \\
0 & 0 & \frac{(d_g+d_m)\mu_g}{d_g\epsilon_m+d_m\mu_g} \\
\end{array}
\right].
\eeq
	
\end{subequations}
	
We can easily find suitable materials as the gyromagnetic medium and the metal slab to 
construct a GHM described above. For example, yttrium iron garnet (YIG) under an applied magnetic 
field of 1600 Gauss has the effective relative permittivity $\epsilon_g=15$ and permeability $\mu_g=1.12$ 
and $\gamma=0.124$ at 1.94 THz~\cite{QAH_Wangzhen_theory}, and thus can be used as the gyromagnetic medium. InSb has the 
effective relative permittivity $\epsilon_m$ = -10.78 at 1.94 THz~\cite{Wu2014} and thus can serve as the metal slab. 
In the bilayer multilayer, the thicknesses of the gyromagnetic medium and metal layers are taken to be 
the same and much less than 1.546$\times$ $10^{-4}$ m (wavelength of 1.94 THz). In this case, 
the effective relative permittivity and permeability tensors can be written as

\begin{subequations}
\beq
\epsilon_{eff}=\left[
\begin{array}{ccc}
2.11 & 0 & 0 \\
0 & 2.11 & 0 \\
0 & 0 & -76.73 \\
\end{array}
\right], 
\eeq

\beq
\mu_{eff}=\left[
\begin{array}{ccc}
1.06 & 0.062i & 0 \\
-0.062i & 1.06 & 0 \\
0 & 0 & 1.06 \\
\end{array}
\right].
\eeq
\end{subequations}

The EFS dispersions obtained using the effective parameters are also displayed in Fig. 8. Clearly, 
the shape of the EFS dispersions is similar to that of the EFS presented in the last section, 
and furthermore, the topological properties such as the Chern numbers of the bands and the topological nature 
of the band gaps are identical to those shown in the last section.

\begin{figure}[htbp]
\includegraphics[width=3.2in]{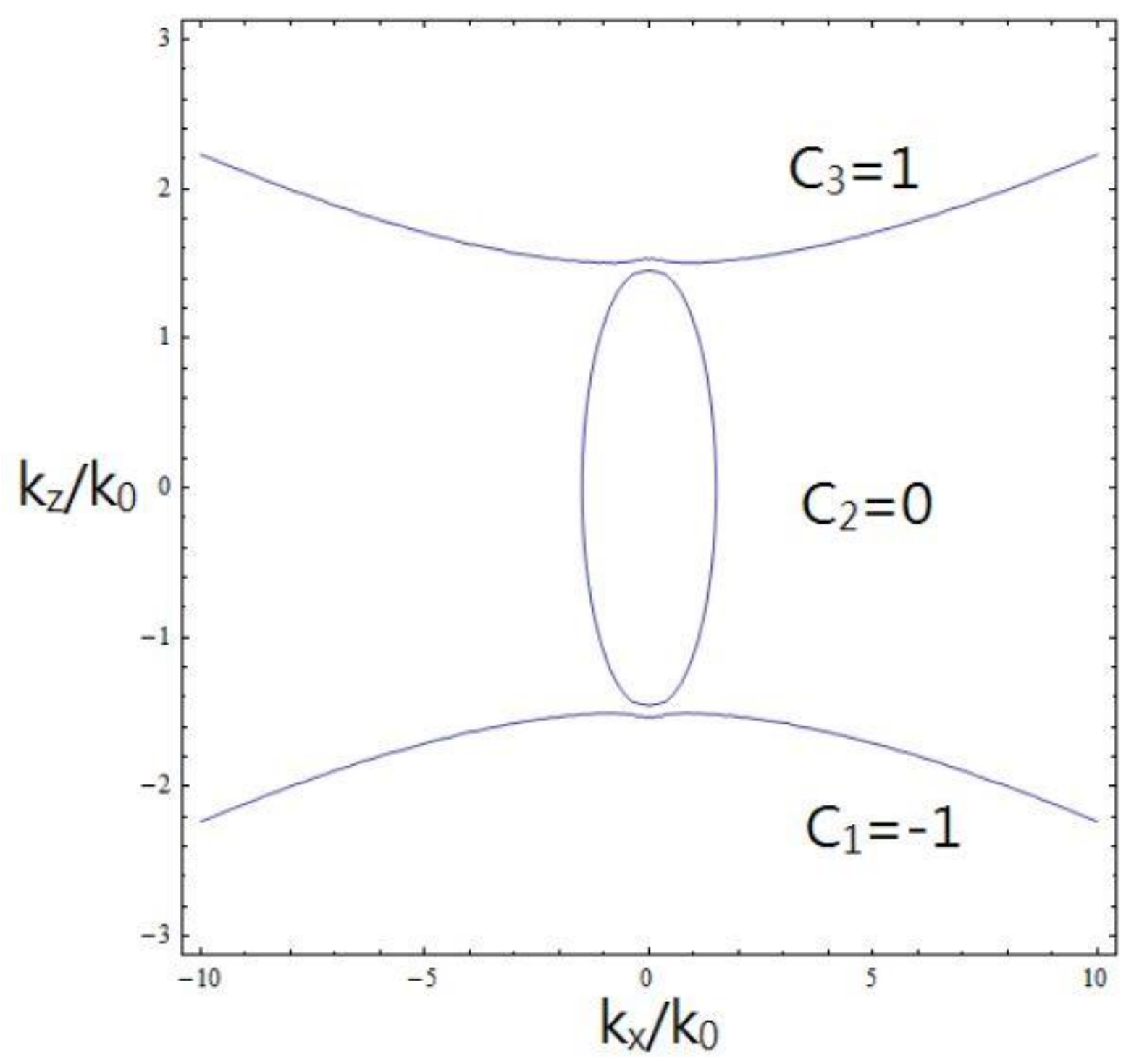}
\caption{Calculated EFS of the bilayer superlattice shown in Fig. 7
made of a YIG slab as the gyromagnetic medium and an InSb layer as the metal slab.}
\end{figure}

\begin{figure}[htbp]
\includegraphics[width=3.2in]{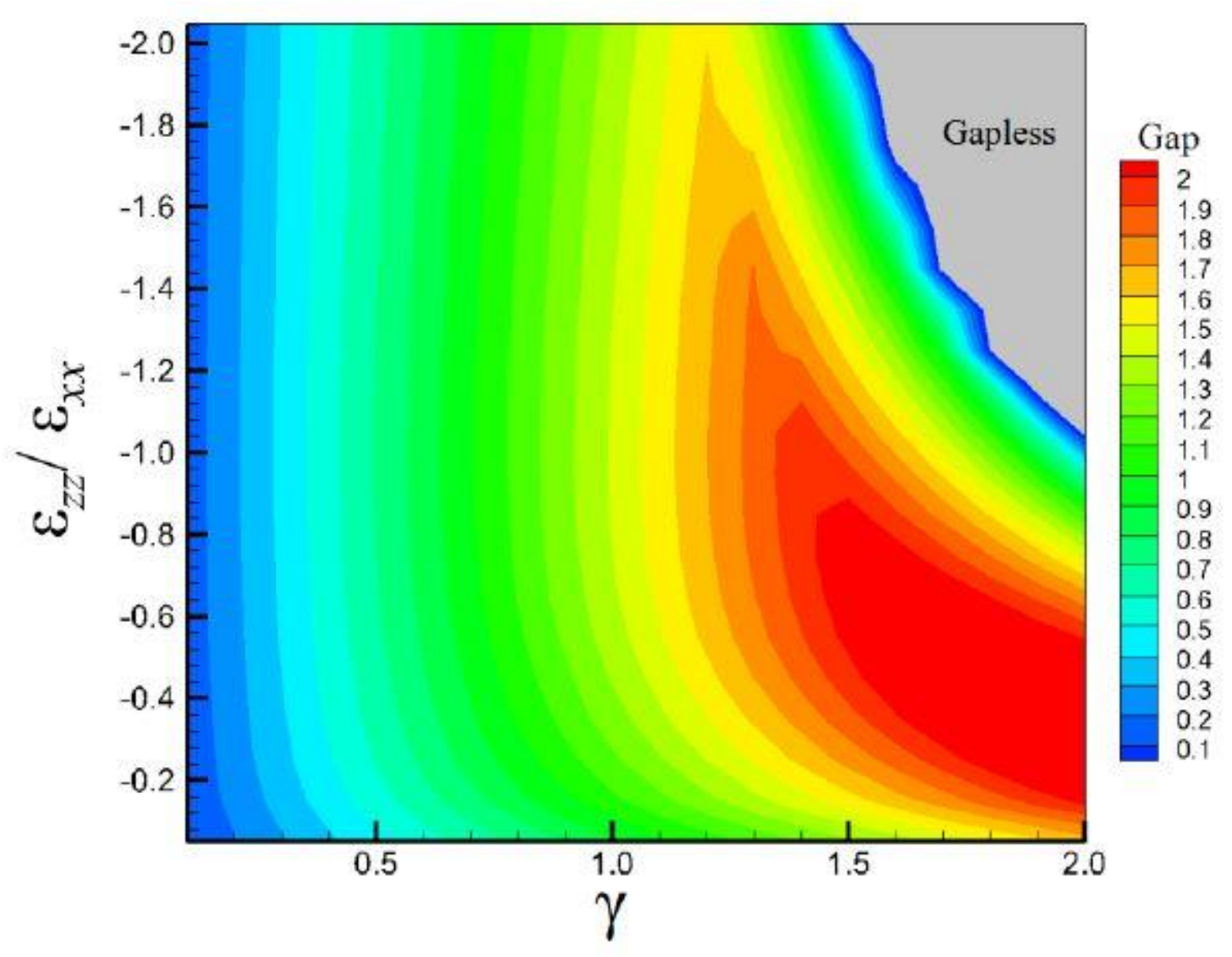}
\caption{Calculated band gap map on the $\gamma-(\epsilon_{zz}/\epsilon_{xx})$ plane of the bilayer superlattice 
shown in Fig. 7, which is made of a YIG slab as the gyromagnetic medium and an InSb layer as the metal slab.}
\end{figure}

To see the robustness of the topological band gap against the variations of the material parameters, 
we show the calculated band gap as a function of both $\gamma$ and $\epsilon_{zz}/\epsilon_{xx}$ in Fig. 9. 
Clearly, the band gap is significant $(>0.5)$ in a wide range of the parameters. For example, 
when $\epsilon_{zz}/\epsilon_{xx}$ is in between -0.1 and -1.0, the band gap is significant for $\gamma$ 
being from 0.4 to 2.0. Interestingly, Fig. 9 shows that for $\epsilon_{zz}/\epsilon_{xx}$ is in 
between -1.0 and -2.0, one tunes the system from the gapless to the Chern insulating state 
and then back to gapless state.
Finally, it should be pointed out that the present photonic Chern insulator would be much simpler 
to design and fabricate than many other photonic insulators. For example, the photonic Floquet 
topological insulators reported in Ref. [\onlinecite{floquetrechtsman2013}] are composed of an array of 
micrometer-scale helical waveguides. The photonic chiral hyperbolic topological insulators proposed 
in Ref. [\onlinecite{chiralhypermeta}] would be based on an array of hyperbolic coils which serve as chiral resonators. 
In contrast, as discussed above, the photonic Chern insulators would consist of a simple YIG/InSb bilayer 
superlattice, and thus would be much easier to fabricate.

\section{Discussion and conclusions}

In the QAH effect, vacuum is an insulator for electrons and thus the chiral edge states are localized 
at the surface of the Chern insulator. In contrast, vacuum to photons is often like a free electron 
metal and thus light propagation on the unidirectional edge modes would generally 
suffer from radiation loss. Consequently, a metal film such as copper is usually inserted between air 
and the topological photonic insulator to suppress the radiation leakage\cite{QAH_Wangzhen_exp,QAH_poo2013,QAH_largeChern_exp}. 
However, metals like copper could incur significant Ohmic loss. In strong contrast, the chiral edge modes 
in our GMHs are not only reflection-free but also non-radiative, as demonstrated 
by our finite element electromagnetic simulations (see Figs. 4 and 5). This is simply 
because the chiral edge dispersions are located outside light cone 
[see Fig. 4(b)]. This is an important advantage of the GMHs 
over the gyromagnetic photonic crystals\cite{QAH_Wangzhen_exp,QAH_poo2013}.

In conclusion, by both theoretical analysis and electromagnetic simulations,
we have demonstrated that gyromagnetic hyperbolic metamaterials (GHM)
are photonic Chern insulators with fascinating properties.
We further show that the large topogical band gaps with gap Chern number of one in these metamaterials, 
result from the simultaneous presence of the hyperbolicity and also the gyromagnetic effect,
which breaks the time-reversal symmetry and thus gives rise to nonzero Berry curvatures on the EFSs.
Remarkably, unlike many other photonic Chern insulators, the GHM Chern insulators possess
non-radiative chiral edge modes on their surfaces, and thus allow to fabricate unidirectional waveguides
without cladding metals which generally incurr considerable Ohmic loss.
Furthermore, the photonic edge states in the proposed Chern insulators are robust against disorder
on a wide range of length scales, in strong contrast to crystalline topological insulators,
and the light flow direction on the surface of the Chern insulators can be easily
flipped by switching the direction of an applied magnetic field. 
We also uncover negative refraction of the topological surface wave at the boundary between
the GHMs with the opposite signs of gyromagnetic parameters. Finally, we show that
compared with other photonic topological materials such as chiral hyperbolic materials\cite{chiralhypermeta}, the present
GHM Chern insulators can be much easier to fabricate.

\begin{acknowledgments}
This work is supported by the Ministry of Science and Technology, the National Center for Theoretical Sciences 
and the Thematic Research Program (AS-TP-106-M07) of the Academia Sinica of the R. O. C. (Taiwan).
\end{acknowledgments}


\end{document}